\documentclass[twocolumn,showpacs,floatfix,aps,prl]{revtex4}

\usepackage[dvips]{graphicx}
\usepackage{color}
\usepackage{dcolumn}
\usepackage{bm}
\usepackage{amsmath, amssymb, mathrsfs}
\usepackage{tikz}
\usetikzlibrary{calc,patterns,decorations.pathmorphing,decorations.markings}

\newcommand{\be}{\begin{equation}}
\newcommand{\ee}{\end{equation}}
\newcommand{\bmul}{\begin{multline}}
\newcommand{\emul}{\end{multline}}
\newcommand{\bea}{\begin{eqnarray}}
\newcommand{\eea}{\end{eqnarray}}
\newcommand{\rr}{\mathbf{r}}
\newcommand{\kk}{\mathbf{k}}
\newcommand{\qq}{\mathbf{q}}
\newcommand{\vv}{\mathbf{v}}

\newcommand{\pp}{\mathbf{p}}
\newcommand{\zero}{\mathbf{0}}

\newcommand{\eee}{\mathrm{e}}
\newcommand{\ii}{\mathrm{i}}
\newcommand{\dd}{\mathrm{d}}
\newcommand{\g}[1]{``#1"}
\newcommand{\bb}[1]{\left( #1 \right)}
\newcommand{\bbcro}[1]{\left[ #1 \right]}
\def\yvan{}
\def\gilles{}

\begin{document}

\title{Landau phonon-roton theory revisited for superfluid helium 4 and Fermi gases}

\author{Yvan Castin and Alice Sinatra}
\affiliation{Laboratoire Kastler Brossel, ENS-PSL, CNRS, Sorbonne Universit\'es, Coll\`ege de France, 75005 Paris, France}

\author{Hadrien Kurkjian}
\affiliation{TQC, Universiteit Antwerpen, Universiteitsplein 1, B-2610 Antwerp, Belgium}


\begin{abstract}
Liquid helium and spin-1/2 cold-atom Fermi gases both exhibit in their superfluid phase two distinct types of excitations, gapless phonons and gapped rotons or fermionic pair-breaking 
excitations. In the long wavelength limit, revising and extending Landau and Khalatnikov's theory initially developed for helium [ZhETF {\bf 19}, 637 (1949)],
we obtain universal expressions for three- and four-body couplings among these two types of excitations. We calculate the corresponding phonon damping rates at low
temperature and compare them to those of a pure phononic origin {\yvan in high-pressure liquid helium and in strongly interacting Fermi gases}, paving the way to experimental observations.
\end{abstract}


\maketitle

\noindent{\sl Introduction --}
Homogeneous superfluids with short-range interactions exhibit, at sufficiently low temperature, phononic excitations {\gilles $\phi$ } as the only microscopic 
degrees of freedom. In this universal limit, all superfluids of this type reduce to a weakly interacting phonon gas {\yvan with a quasilinear dispersion relation}, 
irrespective of the statistics of the underlying particles and of their interaction strength.
Phonon damping then only depends on the dispersion relation close to zero wavenumber (namely, its slope
and third derivative) and on the phonon nonlinear coupling, deduced solely from the system equation of state through Landau-Khalatnikov quantum hydrodynamics
\cite{LKhydroquant}.

In experiments, however, temperatures are not always low enough to make the dynamics purely phononic. Other elementary excitations can enrich the problem,
such as {\yvan spinless bosonic} rotons in liquid helium 4 and {\yvan spinful} fermionic BCS-type pair-breaking excitations in spin-1/2 cold-atom Fermi gases. These excitations, denoted here as $\gamma$-quasiparticles,
exhibit in both cases an energy gap $\Delta>0$. Remarkably, as shown by Landau and Khalatnikov \cite{LKhydroquant}, the phonon-roton coupling, and more generally phonon coupling to
all gapped excitations as we shall see, depend to leading order in temperature only on a few parameters of the dispersion relation of the $\gamma$-quasiparticles,
namely the value of the minimum $\Delta$ and its location $k_0$ in wavenumber space, their derivatives with respect to density, and the effective mass $m_*$
close to $k=k_0$. We have discovered however that the $\phi-\gamma$ coupling of Ref.\cite{LKhydroquant} is not exact, a fact apparently unnoticed in the literature.
Our goal here is to complete the result of Ref.\cite{LKhydroquant}, and to quantitatively obtain phonon damping rates due to the $\phi-\gamma$ coupling as functions of
temperature, a nontrivial task in the considered strongly interacting systems.
We restrict to the collisionless regime $\omega_\qq\tau_\gamma\gg 1$ and $\omega_\qq\tau_\phi\gg 1$, where $\omega_\qq$ is the angular eigenfrequency of the considered
phonon mode of wavevector $\qq$, and $\tau_\gamma$ ($\tau_\phi$) is a typical collision time of thermal $\gamma$-quasiparticles (thermal phonons). 
An extension to the hydrodynamic regime $\omega_\qq\tau_\gamma\lesssim 1$ or $\omega_\qq\tau_\phi\lesssim 1$ may be obtained from kinetic equations \cite{Chernikova}.
An experimental test of our results seems nowadays at hand, either in liquid helium 4, extending the recent work of Ref.\cite{Fok}, or in homogeneous cold Fermi gases,
which the breakthrough of flat-bottom traps \cite{boite} allows one to prepare \cite{boite_fermions} 
{\gilles and to acoustically excite by spatio-temporally modulated laser-induced optical potentials \cite{EPL,MZP}.}

\noindent{\sl Landau-Khalatnikov revisited --}
We recall the reasoning of Ref.\cite{LKhydroquant} to get the phonon-roton coupling in liquid helium 4, extending it to the phonon-fermionic quasiparticle coupling
in unpolarised spin-1/2 Fermi gases. We first treat in first quantisation the case of a single roton or fermionic excitation, considered as a $\gamma$-quasiparticle of position $\rr$, momentum $\pp$ and spin $s=0$ or $s=1/2$. In a homogeneous superfluid of density $\rho$, its Hamiltonian is given by $\epsilon(\pp,\rho)$,
an isotropic function of $\pp$ such that $p \mapsto \epsilon(\pp,\rho)$ is the $\gamma$-quasiparticle dispersion relation.
In presence of acoustic waves (phonons), the superfluid acquires position-dependent density $\rho(\rr)$ and velocity $\vv(\rr)$.
For a phonon wavelength large compared to the $\gamma$-quasiparticle coherence length {\yvan \cite{CCT}},  
{\gilles here} its thermal 
wavelength $({2\pi\hbar^2}/{m_* k_B T})^{1/2}$ \footnote{\yvan\gilles For a thermal phonon wavenumber, this requires $k_B T\ll m_*c^2$, a meaningful condition even in the BEC limit of Fermi gases where $k_0=0$.},
and {\yvan for} a phonon angular frequency small compared to the $\gamma$-quasiparticle \g{internal} energy $\Delta$,
we can write the $\gamma$-quasiparticle Hamiltonian in the local density approximation {\yvan \cite{Thomas,Fermi}}:
\be
\mathcal{H} = \epsilon(\pp,\rho(\rr)) + \pp\cdot \vv(\rr)
\label{eq:hrot}
\ee
The last term is a Doppler effect reflecting the energy difference in the lab frame and in the frame moving with the superfluid.
For a weak phononic perturbation of the superfluid, we expand the Hamiltonian to second order in density fluctuations $\delta\rho(\rr)=\rho(\rr)-\rho$:
\be
\mathcal{H}\simeq \epsilon(\pp,\rho) + \partial_\rho \epsilon(\pp,\rho) \delta\rho(\rr) +\pp\cdot \vv(\rr)
+\frac{1}{2} \partial_\rho^2 \epsilon(\pp,\rho) \delta\rho^2(\rr)
\label{eq:hrotdev}
\ee
not paying attention yet to the noncommutation of $\rr$ and $\pp$. Phonons are bosonic quasiparticles connected to the expansion of $\delta\rho(\rr)$ and $\vv(\rr)$
on eigenmodes of the quantum-hydrodynamic equations linearised around the homogeneous solution at rest in the quantisation volume $\mathcal{V}$:
\be
\begin{pmatrix}\delta\rho(\rr) \\ \vv(\rr) \end{pmatrix}
= \frac{1}{\mathcal{V}^{1/2}} \sum_{\qq\ne\zero} \left[\binom{\rho_q}{\vv_\qq} \hat{b}_\qq+
\binom{\rho_q}{-\vv_\qq} \hat{b}_{-\qq}^\dagger\right] \eee^{\ii \qq\cdot\rr}
\ee
with modal amplitudes $\rho_q=[{\hbar\rho q}/({2mc})]^{1/2}$ and $\vv_\qq = [{\hbar c}/({2m\rho q})]^{1/2} \qq$, $m$ being the mass of a superfluid particle and $c$ the sound velocity.
The annihilation and creation operators $\hat{b}_\qq$ and $\hat{b}_\qq^\dagger$ of a phonon of wavevector $\qq$ and energy $\hbar\omega_\qq=\hbar cq$ obey usual commutation relations $[\hat{b}_\qq,\hat{b}_{\qq'}^\dagger] = \delta_{\qq,\qq'}$.  

For an arbitrary number of $\gamma$-quasiparticles, we switch to second quantisation and rewrite Eq.(\ref{eq:hrotdev}) as 
\begin{multline}
\label{eq:hrotdevsec}
\hat{H} = \sum_{\kk,\sigma} \epsilon_\kk \hat{\gamma}_{\kk\sigma}^\dagger \hat{\gamma}_{\kk\sigma} 
+ \!\sum_{\kk,\kk',\qq,\sigma}\!\! \frac{\mathcal{A}_1(\kk,\qq;\kk')}{\mathcal{V}^{1/2}}
(\hat{\gamma}_{\kk'\sigma}^\dagger \hat{\gamma}_{\kk\sigma} \hat{b}_\qq + \mbox{h.c.}) \\ \times \delta_{\kk+\qq,\kk'} 
+\!\!\!\sum_{\kk,\kk',\qq,\qq',\sigma}\!\!\! \frac{\mathcal{A}_2(\kk,\qq ; \kk',\qq')}{\mathcal{V}} 
\hat{\gamma}_{\kk'\sigma}^\dagger\hat{\gamma}_{\kk\sigma} \delta_{\kk+\qq,\kk'+\qq'} \\
\times [\hat{b}_{\qq'}^\dagger \hat{b}_\qq + \frac{1}{2} (\hat{b}_{-\qq'}\hat{b}_\qq + \mbox{h.c.})]
\end{multline}
where $\hat{\gamma}_{\kk\sigma}$ and $\hat{\gamma}_{\kk\sigma}^\dagger$ are bosonic (rotons, $s=0$, $\sigma=0$) or fermionic ($s=1/2$, $\sigma=\uparrow,\downarrow$)
annihilation and creation operators of a $\gamma$-quasiparticle of wavevector $\kk=\pp/\hbar$ in spin component $\sigma$, obeying usual commutation or
anticommutation relations.
The first sum in the right-hand side of Eq.(\ref{eq:hrotdevsec}) gives the $\gamma$-quasiparticle energy in the unperturbed superfluid, with
$\epsilon_\kk \equiv \epsilon(\hbar\kk,\rho)$. The second sum, originating from the Doppler term and the term linear in $\delta\rho$ in Eq.(\ref{eq:hrotdev}),
describes absorption or emission of a phonon by a $\gamma$-quasiparticle, characterised by the amplitude 
\be
\label{eq:A1}
\mathcal{A}_1(\kk,\qq;\kk')=\rho_q \frac{\partial_\rho \epsilon_\kk + \partial_\rho \epsilon_{\kk'}}{2} 
+\vv_\qq \cdot \frac{\hbar\kk+\hbar\kk'}{2}
\ee
where $\qq$, $\kk$ and $\kk'$ are the wavevectors of the incoming phonon and the incoming and outgoing $\gamma$-quasiparticles.
Eq.(\ref{eq:A1}) is invariant under exchange of $\kk$ and $\kk'$. This results from symmetrisation of the various terms, in the form
$[f(\pp) \eee^{\ii\qq\cdot\rr}+\eee^{\ii\qq\cdot\rr} f(\pp)]/2$ with $\rr$ and $\pp$ canonically conjugated operators, 
ensuring that the correct form of Eq.(\ref{eq:hrotdev})
is hermitian. The third sum in Eq.(\ref{eq:hrotdevsec}), originating from the terms
quadratic in $\delta\rho$ in Eq.(\ref{eq:hrotdev}), describes direct scattering of a phonon on a $\gamma$-quasiparticle,
with the symmetrised amplitude
\be
\label{eq:A2}
\mathcal{A}_2(\kk,\qq;\kk',\qq')=\rho_q \rho_{q'} \frac{\partial^2_\rho\epsilon_\kk + \partial^2_\rho\epsilon_{\kk'}}{2}
\ee
where the primed wavevectors are the ones of emerging quasiparticles. It also describes negligible two-phonon absorption and emission.
The effective amplitude for $\phi-\gamma$ scattering is obtained by adding the contributions of the direct process {\yvan (terms of $\hat{H}$ quadratic
in $\hat{b}$)}, and of the
absorption-emission or emission-absorption process {\yvan (terms linear in $\hat{b}$)} treated to second order in perturbation theory \cite{LKhydroquant}:

\begin{multline}
\label{eq:A2eff}
\mathcal{A}_2^{\rm eff}(\kk,\qq;\kk',\qq') = \!\!\!\!
{\vcenter{\hbox{
\begin{tikzpicture}[scale=0.25]
\draw[decoration={snake,pre length=0.0cm,post length=0.0cm,segment length=5, amplitude=0.05cm},decorate] (-2,1.5) -- node [left=0.1cm] {\scriptsize$\qq$}(0,0);
\draw[decoration={snake,pre length=0.0cm,post length=0.0cm,segment length=5, amplitude=0.05cm},decorate] (0,0) -- node [right=0.1cm] {\scriptsize$\qq'$}(2,-1.5);
\draw[-] (-2,-1.5) --   node[left]{\scriptsize$\kk$} (0,0);
\draw[-] (0,0) --   node[right]{\scriptsize$\kk'$} (2,1.5);
\end{tikzpicture}
}}\!\!\!+\!\!\!\vcenter{\hbox{
\begin{tikzpicture}[scale=0.25]
\draw[decoration={snake,pre length=0.0cm,post length=0.0cm,segment length=5, amplitude=0.05cm},decorate] (-1.5,1.5) -- node [left] {\scriptsize$\qq$}(0,0);
\draw[-] (-1.5,-1.5) --   node[left]{\scriptsize$\kk$} (0,0);
\draw[-] (0,0) --   (1.5,0);
\draw[-] (1.5,0) --  node[right]{\scriptsize$\kk'$} (3,1.5);
\draw[decoration={snake,pre length=0.0cm,post length=0.0cm,segment length=5, amplitude=0.05cm},decorate] (1.5,0) -- node [right] {\scriptsize$\qq'$}(3,-1.5);
\end{tikzpicture}}}
+
\!\!\!\!
\vcenter{\hbox{
\begin{tikzpicture}[scale=0.25]
\draw[decoration={snake,pre length=0.0cm,post length=0.0cm,segment length=5, amplitude=0.05cm},decorate] (0,1.5) -- node [left] {\scriptsize$\qq$}(1.5,0);
\draw[-] (-1.5,-1.5) --   node[left]{\scriptsize$\kk$} (0,0);
\draw[-] (0,0) --   (1.5,0);
\draw[-] (1.5,0) --  node[right]{\scriptsize$\kk'$} (3,1.5);
\draw[decoration={snake,pre length=0.0cm,post length=0.0cm,segment length=5, amplitude=0.05cm},decorate] (0,0) -- node [right] {\scriptsize$\qq'$}(1.5,-1.5);
\end{tikzpicture}}}}
\\
=\mathcal{A}_2(\kk,\qq;\kk',\qq')
+\frac{\mathcal{A}_1(\kk,\qq;\kk+\qq) \mathcal{A}_1(\kk',\qq';\kk'+\qq')}{\hbar\omega_\qq + \epsilon_\kk-\epsilon_{\kk+\qq}} \\
+\frac{\mathcal{A}_1(\kk-\qq',\qq';\kk) \mathcal{A}_1(\kk-\qq',\qq;\kk')}{\epsilon_\kk-\hbar\omega_{\qq'}-\epsilon_{\kk-\qq'}}
\end{multline}
{\yvan where in the second (third) term the $\gamma$-quasiparticle first absorbs phonon $\qq$ (emits phonon $\qq'$) 
then emits phonon $\qq'$ (absorbs phonon $\qq$).}
Up to this point this agrees with Ref.\cite{LKhydroquant}, except 
that the first derivative $\partial_\rho\Delta$ in Eq.(\ref{eq:A1}), thought to be anomalously small in low-pressure helium, was neglected in Ref.\cite{LKhydroquant}.

Eq.~(\ref{eq:A2eff}), issued from a local density approximation, holds to leading order in a low-energy limit.
We then take the $T\to 0$ limit with scaling laws
\be
q\approx T, \ \ \ \ k-k_0 \approx T^{1/2}
\label{eq:loiech}
\ee
reflecting the fact that the thermal energy of a phonon is $\hbar cq \approx k_B T$ and the effective kinetic energy of a 
$\gamma$-quasiparticle, that admits the expansion 
\be
\label{eq:ekdev}
\epsilon_\kk -\Delta \underset{k\to k_0}{=} \frac{\hbar^2 (k-k_0)^2}{2 m_*} + O(k-k_0)^3
\ee
is also $\approx k_B T$.  The coupling amplitudes $\mathcal{A}_1$ and energy denominators in Eq.(\ref{eq:A2eff}) 
must be expanded up to relative corrections of order $T$
\footnote{\gilles One expands to order $T^{3/2}$ for $\mathcal{A}_1$ and $T^2$ for energy denominators, 
$q'$ being deduced from $\qq$, $\kk$ and $\qq'/q'$ by energy conservation,
$q-q'=\frac{\hbar(k-k_0)q(u-u')}{m_*c}[1+\frac{\hbar(k-k_0)u'}{m_*c}] +\frac{\hbar q^2(u-u')^2}{2m_*c}+O(T^{5/2})$ with $u$ and $u'$ defined below Eq.(\ref{eq:A2effequiv}).}.
On the contrary, it suffices to expand $\mathcal{A}_2$ to leading order $T$ in temperature.
We hence get our main result, the effective coupling amplitude of the 
$\phi-\gamma$ {\gilles scattering} to leading order in temperature:
\begin{multline}
\label{eq:A2effequiv}
\mathcal{A}_2^{\rm eff}(\kk,\qq;\kk',\qq')  \underset{T\to 0}{\sim} \frac{\hbar q}{mc\rho}
\Bigg\{\frac{1}{2} \rho^2 \Delta'' + \frac{(\hbar \rho k_0')^2}{2m_*} 
+ \frac{\hbar^2 k_0^2}{2m_*} \\ \times \Bigg\{\!\!\left(\frac{\rho\Delta'}{\hbar c k_0}\right)^2\!\! u u' + \frac{\rho\Delta'}{\hbar c k_0}
\left[(u+u')\left(u u' -\frac{\rho k_0'}{k_0}\right) + \frac{2m_* c}{\hbar k_0} w\right]\\
+ \frac{m_* c}{\hbar k_0} (u+u') w + u^2 u'^2 -\frac{\rho k_0'}{k_0} (u^2+u'^2) \Bigg\}\Bigg\}
\end{multline}
Here $\Delta'$, $k_0'$, $\Delta''$ are first and second derivatives of $\Delta$ and $k_0$ with respect to $\rho$;
$u=\frac{\qq\cdot\kk}{qk}$, $u'=\frac{\qq'\cdot\kk}{q'k}$, $w=\frac{\qq\cdot\qq'}{qq'}$ are cosines of the angles between $\kk$, $\qq$ and $\qq'$; 
our results hold for $k_0=0$ provided the limit $k_0\to0$ is taken in Eq.(\ref{eq:A2effequiv}).
In {\yvan Eq.(3.17) of} Ref.\cite{LKhydroquant}, the $\Delta'$ terms were neglected as said, but the last term in Eq.(\ref{eq:A2effequiv}), 
with the factor $\rho k_0'/k_0$, was simply forgotten.

{\bf Addition in the corrected-augmented version:} {\it Expression \eqref{eq:A2eff} of the coupling amplitude is incomplete because it does not take into account the interactions among phonons. This omission affects expression \eqref{eq:A2effequiv} of the effective amplitude, the angular integral in note [47] and the dashed lines of Figs.~\ref{fig:helium} and \ref{fig:gaz}. The erratum \cite{erratum} that corrects this omission is reproduced here in appendix, see in particular Eqs.~\eqref{eq:A2efferr} and \eqref{eq:A2effequiv2}, and it is supplemented by a verification of the final result using a microscopic approach based on Bogoliubov theory with an arbitrary short-range interaction potential.} -- {\bf End of addition.} \\

\noindent{\sl Damping rates --} 
A straightforward application of Eq.(\ref{eq:A2effequiv}) is a Fermi-golden-rule calculation of the damping rate $\Gamma_\qq^{\gilles\rm scat}$ of phonons $\qq$ due to scattering on $\gamma$-quasiparticles. 
The $\gamma$-quasiparticles are in thermal equilibrium with Bose or Fermi mean occupation numbers 
$\bar{n}_{\gamma,\kk}=[\exp(\epsilon_\kk/k_B T)-(-1)^{2s}]^{-1}$. So are phonons in modes $\qq' \neq \qq$,
with Bose occupation numbers $\bar{n}_{b,\qq'}=[\exp(\hbar\omega_{\qq'}/k_B T)-1]^{-1}$; mode $\qq$ is initially excited {\yvan (e.g.\ by a sound wave)}
with an arbitrary number $n_{b,\qq}$ of phonons. By including both loss $\qq+\kk\to\qq'+\kk'$ and gain $\qq'+\kk'\to\qq+\kk$ processes
\footnote{\gilles We also use energy conservation and the relations $1+(-1)^{2s}\bar{n}=\eee^{\epsilon/k_B T} \bar{n}$ to transform 
the difference of the gain and loss quantum statistical factors.}
{\gilles and summing over $\sigma$}, one finds that
$\frac{\dd}{\dd t} n_{b,\qq} = - \Gamma_\qq^{\gilles\rm scat} (n_{b,\qq}-\bar{n}_{b,\qq})$ with
\begin{multline}
\label{eq:gammaelor}
\Gamma_\qq^{\gilles\rm scat} = \frac{2\pi}{\hbar} (2s+1)
\int \frac{\dd^3k \dd^3q'}{(2\pi)^6} \left[\mathcal{A}_2^{\rm eff}(\kk,\qq;\kk',\qq')\right]^2 \\
\times \delta(\epsilon_\kk+\hbar\omega_\qq-\epsilon_{\kk'}-\hbar\omega_{\qq'}) \frac{\bar{n}_{b,\qq'} \bar{n}_{\gamma,\kk'} 
[1+ (-1)^{2s}\bar{n}_{\gamma,\kk}]}{\bar{n}_{b,\qq}}
\end{multline}
and $\kk'=\kk+\qq-\qq'$. As our low-energy theory only holds for $k_B T\ll \Delta$, the gas of $\gamma$-quasiparticles is nondegenerate, and
$\bar{n}_{\gamma,\kk}\simeq \exp(-\epsilon_\kk/k_B T) \ll 1$ in Eq.(\ref{eq:gammaelor}). By taking the $T\to 0$ limit at fixed 
$\hbar c q/k_B T$ and setting $\mathcal{A}_2^{\rm eff}=\frac{\hbar\omega_{\yvan\qq}}{\rho} f$, 
where the dimensionless quantity $f$ only depends on angle cosines, we obtain the equivalent 
\be
\label{eq:gamelequiv}
\hbar\Gamma_\qq^{\gilles\rm scat} \underset{T\to 0}{\sim}(2s+1)\frac{\eee^{-\Delta/k_B T}}{(2\pi)^{9/2}} \frac{k_0^2 q^4 c}{\rho^2} (m_* k_B T)^{1/2} I
\ee
with $I\!=\!\int\dd^2\Omega_\kk\int\dd^2\Omega_{\qq'}f^2(u,u',w)$ an integral over solid angles of direction $\kk$ and $\qq'$ 
\footnote{\gilles $I/(4\pi)^2=
(\frac{\hbar^2 k_0^2}{2m_*mc^2})^2[\frac{1}{25}-\frac{4\alpha}{15}+\frac{28}{45}\alpha^2+\frac{2\beta^2}{9}
+A(\frac{2}{9}-\frac{4\alpha}{3})+A^2+4\beta B(\frac{1}{15}-\frac{\alpha}{9})
+B^2(\frac{2}{15}-\frac{4\alpha}{9}+\frac{2\alpha^2}{3}+\frac{4\beta^2}{3})+\frac{4\beta}{9}B^3+\frac{B^4}{9}]$,
with $\alpha=\frac{\rho k_0'}{k_0}$, $\beta=\frac{m_*c}{\hbar k_0}$, $A=\frac{m_*\rho^2\Delta''}{(\hbar k_0)^2} +\alpha^2$, $B=\frac{\rho \Delta'}{\hbar c k_0}$.}.

One proceeds similarly for the calculation of the damping rate $\Gamma_\qq^{\gilles\mbox{\scriptsize a-e}}$ of phonons $\qq$ due to absorption  $\qq+\kk\to \kk'$ or emission $\kk'\to\qq+\kk$ {\gilles processes} by thermal equilibrium $\gamma$-quasiparticles. We obtain 
\begin{multline}
\label{eq:gammainelor}
\Gamma_\qq^{\gilles\mbox{\scriptsize a-e}} = \frac{2\pi}{\hbar} (2s+1)\int \frac{\dd^3k}{(2\pi)^3} [\mathcal{A}_1(\kk,\qq;\kk')]^2 \\  
\times \delta(\hbar\omega_\qq+\epsilon_\kk-\epsilon_{\kk'}) (\bar{n}_{\gamma,\kk}-\bar{n}_{\gamma,\kk'})
\end{multline}
with $\kk'=\kk+\qq$. Low degeneracy of the $\gamma$-quasiparticles and energy conservation
allow us to write $\bar{n}_{\gamma,\kk}-\bar{n}_{\gamma,\kk'}\simeq \exp(-\epsilon_\kk/k_B T)/(1+\bar{n}_{b,\qq})$.
Energy conservation leads here to a scaling on $k$ different from
Eq.(\ref{eq:loiech}) as it forces $k$ to be
at a nonzero distance from $k_0$, even in the low-phonon-energy limit: 
When $q\to 0$ at fixed $\kk$, the Dirac delta in Eq.(\ref{eq:gammainelor}) becomes
\be
\label{eq:Dirac}
\delta(\hbar\omega_\qq+\epsilon_\kk-\epsilon_{\kk'}) \underset{q\to 0}{\sim} (\hbar c q)^{-1} 
\delta\left(1-u\frac{\frac{\dd\epsilon_k}{\dd k}}{\hbar c}\right)
\ee
and imposes that the group velocity $\frac{1}{\hbar} \frac{\dd\epsilon_k}{\dd k}$ of the incoming 
$\gamma$-quasiparticle is larger in absolute value than that, $c$, of the phonons. 
This condition, reminiscent of Landau's criterion, restricts wavenumber $k$ to a {\gilles domain $D$ not containing $k_0$}. 
In the low-$q$ limit, {\gilles that is for $q$ much smaller than the $k$} significantly contributing to Eq.(\ref{eq:gammainelor}), 
but with no constraint on the ratio $\hbar c q/k_B T$,
we write $\mathcal{A}_1$ in Eq.(\ref{eq:A1}) to leading order $q^{1/2}$ in $q$, and integrate over the direction of $\kk$, to obtain
\bea
\label{eq:gammainelzero}
\!\!\!\!\Gamma_\qq^{\gilles\mbox{\scriptsize a-e}}\!\!\!\!&\simeq&\!\!\!\!\frac{(2s+1)\rho}{4\pi mc}\!\!\int_{D}\frac{\dd k k^2}{|\frac{\dd\epsilon_k}{\dd k}|} \frac{\eee^{-\epsilon_\kk/k_B T}}{1+\bar{n}_{b,\qq}}
\left|\partial_\rho\epsilon_k\!+\!\frac{\hbar^2c^2k}{\rho\frac{\dd\epsilon_k}{\dd k}}\right|^2\\
\!\!\!\!\!\!\!\!&\underset{T\to 0}{\sim} &\!\!\!\!\frac{(2s+1)\rho k_*^2}{4\pi\hbar^2 mc^3} \left|\partial_\rho\epsilon_{k_*}\!+\!\frac{\hbar ck_*}{\rho\eta_*}\right|^2
\!\frac{k_B T\eee^{-\epsilon_{k_*}/k_B T}}{1+\bar{n}_{b,\qq}}
\label{eq:gammainelzeroequiv}
\eea
Eq.(\ref{eq:gammainelzeroequiv}) is an equivalent when $T\to 0$ at fixed $\hbar c q/k_B T$; $k_*$ is the element of the border of 
$D$ ($\frac{\dd\epsilon_k}{\dd k}|_{k=k_*}=\eta_* \hbar c$, $\eta_*=\pm$) 
with minimal energy $\epsilon_k$ 
(when more than one of such $k_*$ exists, one has to sum their contributions).
As $\epsilon_{k_*}>\Delta$, the damping rate due to {\gilles scattering} dominates the one {\gilles due to absorption-emission} in the mathematical limit $T\to 0$~;
we shall see however that this is not always so for typical temperatures in current experiments. 

To be complete, we give a low-temperature equivalent of the damping rate of the $\gamma$-quasiparticle $\kk$
due to interaction with thermal phonons. With $k-k_0=O(T^{1/2})$ as in Eq.(\ref{eq:loiech}), we find
$\hbar\Gamma_{\kk}^{\gamma \phi} \sim (\pi I/42) (k_B T)^7/(\hbar c\rho^{1/3})^6$,
where the factor $2s+1$ is gone {\gilles (no summation over $\sigma$ is needed)} 
but $I$ is the same angular integral as in Eq.(\ref{eq:gamelequiv}).
{\gilles Here scattering dominates} \footnote{\gilles At low $T$,
 $k$ is close to $k_0$,  emission $\kk\leftrightarrow \qq+\kk'$ is forbidden by energy conservation, 
and absorption $\kk+\qq\leftrightarrow \kk'$, conserving energy only for 
$q\geq q_*\simeq 2m_*c/\hbar$, is $O(\eee^{-\hbar\omega_{q_*}/k_B T})$. }.
Using $\tau_\gamma\simeq 1/\Gamma_\kk^{\gamma\phi}$, we checked that the figures \ref{fig:helium} and \ref{fig:gaz} below are in the collisionless regime
$\omega_\qq \tau_\gamma\gg 1$.  Similarly, we checked that $\omega_\qq\tau_\phi\gg 1$ on the figures.

\begin{figure}[t]
\begin{center}
\includegraphics[width=0.48\textwidth,clip=]{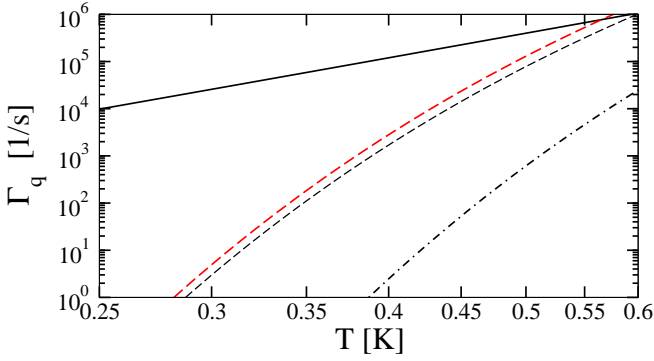}
\caption{Phonon damping rates at angular frequency $\omega_\qq=2\pi\times 165$ GHz ($q=0.3\mbox{\AA}^{-1}$) in liquid ${}^4$He at pressure $P=20$ bar as functions of temperature. Solid line: purely phononic damping $\Gamma_{\phi\phi}$
due to Landau-Khalatnikov four-phonon processes \cite{LKhydroquant,Annalen,EPL}; it depends on the curvature parameter
$\gamma$ defined as $\omega_\qq=cq[1+\frac{\gamma}{8}(\frac{\hbar q}{mc})^2+O(q^4)]$.  
Interpolating measurements of $P\mapsto\gamma(P)$ in Refs.\cite{FosterPRB30_2595,SvensonPRL29_1148} gives $\gamma=-6.9$. Dashed black line/dash-dotted black line: damping due to {\gilles scattering/absorption-emission} by rotons, see Eq.(\ref{eq:gamelequiv})/(\ref{eq:gammainelzero}). Red dashed line: original formula of Ref.\cite{LKhydroquant} for the {\gilles damping rate due to phonon-roton scattering.}
The roton parameters are extracted from their dispersion relation
$k\mapsto\epsilon_\kk$ measured at various pressures \cite{Gibbs1999}:
$\Delta/k_B=7.44$K,  $k_0=2.05 \mbox{\AA}^{-1}$, $m_*/m=0.11$, $\rho k_0'/k_0=0.39$, $\rho\Delta'/\Delta=-1.64$, $\rho^2\Delta''/\Delta=-8.03$, $\rho m_*'/m_*=-4.7$.
In Eq.(\ref{eq:gammainelzero}), parabolic approximation (\ref{eq:ekdev}) is used (hence $\epsilon_{k_*}/\Delta\simeq 1.43$).
The speed of sound $c=346.6$\,m/s, and the Gr\"uneisen parameter $\frac{\dd\ln c}{\dd\ln\rho}=2.274$ entering in $\Gamma_{\phi\phi}$,
are taken from equation of state (A1) of Ref.\cite{Marris2002}. The low values $\frac{\hbar q}{mc}=0.13$ and $\frac{k_B T}{mc^2}<10^{-2}$ 
justify our use of quantum hydrodynamics.
\label{fig:helium}
}
\end{center}
\end{figure}

\begin{figure}[t]
\begin{center}
\includegraphics[width=0.48\textwidth,clip=]{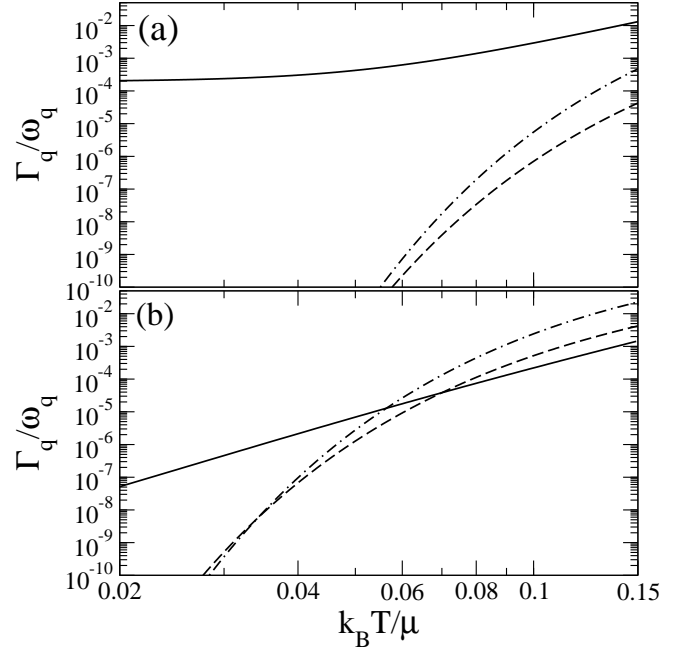}
\caption{Phonon damping rates at wavenumber $q=mc/2\hbar$ in unpolarized homogeneous cold-atom Fermi gases in thermodynamic limit as functions of temperature. (a) At unitarity $a^{-1}=0$, where most parameters of the phonons and fermionic quasiparticles are measured (see text).  (b) On the BCS side $1/k_{\rm F} a=-0.389$, these parameters are estimated in BCS theory
($\mu/\epsilon_{\rm F}\simeq 0.809$, $\Delta/\mu\simeq 0.566$, $\frac{m_*}{m}=\frac{\Delta}{2\mu}$,
$\rho\mu'/\mu\simeq 0.602$, $\rho\Delta'/\Delta\simeq 0.815$, $\rho^2\Delta''/\Delta\simeq -0.209$, 
$\frac{\dd\ln c}{\dd\ln\rho}\simeq 0.303$). 
In both cases the curvature parameter $\gamma$ defined in the caption of Fig.~\ref{fig:helium} is estimated in the RPA \cite{concavrpa}.
Solid line: phonon-phonon (a) Beliaev-Landau damping $\phi\leftrightarrow\phi\phi$ (for $\gamma>0$) as in Eqs.(121,122) of Ref.\cite{Annalen} (independent of $|\gamma|$) and (b) Landau-Khalatnikov damping $\phi\phi\leftrightarrow\phi\phi$ (for $\gamma\simeq -0.30<0$) \cite{Annalen,EPL}. 
Dashed line/dash-dotted line: {\gilles scattering/absorption-emission} phonon-fermionic quasiparticle processes, as in Eq.(\ref{eq:gamelequiv})/(\ref{eq:gammainelzero}). 
In Eq.(\ref{eq:gammainelzero}), we took for $\epsilon_k$ (a) the form proposed in Ref.\cite{Nishida} (hence $\epsilon_{k_*}/\Delta\simeq 1.12$)  and (b) the BCS form (hence $\epsilon_{k_*}/\Delta\simeq 1.14$).
$\mu$ is the $T=0$ gas chemical potential, and the plotted quantities are in fact inverse quality factors. {\gilles Here $k_B T/mc^2>0.03$ in contrast to Fig.\ref{fig:helium} where $k_B T/mc^2<0.01$: cold atoms are effectively farther from the $T\to 0$ limit than liquid helium, hence the inversion of the $\Gamma_{\qq}^{\rm scat}$-$\Gamma_{\qq}^{\mbox{\scriptsize a-e}}$ hierarchy.}}
\label{fig:gaz}
\end{center}
\end{figure}

\noindent{\sl Application to helium --}
Precise measurements of the equation of state (relating $\rho$ to pressure) and of the roton dispersion relation for various pressures were performed in liquid $^4$He at low temperature ($k_B T \ll mc^2, \Delta$). They give access to the parameters $k_0$, $\Delta$, their derivatives and $m_*$. The measured sound velocities agree with the thermodynamic relation $mc^2=\rho \frac{\dd\mu}{\dd\rho}$, where $\mu$ is the zero-temperature chemical potential of the liquid. We plot in Fig.~\ref{fig:helium} the phonon damping rates as functions of temperature, for a fixed angular frequency $\omega_\qq$. At the chosen high pressure, the phonon dispersion relation is concave at low $q$, therefore the Beliaev-Landau \cite{Beliaev,PitaevskiiStringari,Giorgini,mesuresHe,mesuresBose,Annalen} three-phonon process $\phi\leftrightarrow\phi\phi$ is energetically forbidden at low temperature and the Landau-Khalatnikov \cite{LKhydroquant,Annalen,EPL} process $\phi\phi\leftrightarrow\phi\phi$ is dominant. Our high yet experimentally accessible \cite{Lockerbie1974,Dietsche1978} value of $\omega_\qq$ leads to attenuation lengths $2c/\Gamma_\qq$ short enough to be measured in centimetric cells. As visible on Fig.~\ref{fig:helium}, the damping of sound is in fact dominated by four-phonon Landau-Khalatnikov processes up to a temperature $T\simeq 0.6$ K. In this regime one would directly observe this phonon-phonon damping mechanism, which would be a premiere. The sound attenuation measurements of Ref.\cite{Berberich1976} in helium at 23 bars and $\omega_\qq=2\pi\times
1.1$ GHz are indeed limited to $T>0.8$ K where damping is still dominated by the rotons.

\noindent{\sl Application to fermions --}
In cold-atom Fermi gases, interactions occur in $s$-wave between opposite-spin atoms. Of negligible range, they are characterized by the scattering length $a$ tunable by Feshbach resonance \cite{Thomas2002,Salomon2003,Grimm2004b,Ketterle2004,Salomon2010,Zwierlein2012}.

Precise measurements of the fermionic excitation parameters $k_0$ and $\Delta$ were performed at unitarity $a^{-1}=0$ \cite{KetterleGap}. Due to the unitary-gas scale invariance \cite{JasonHo,Zwerger,Zwergerlivre}, $k_0$ is proportional to the Fermi wavenumber $k_{\rm F}=(3\pi^2\rho)^{1/3}$, $k_0\simeq 0.92 k_{\rm F}$ \cite{KetterleGap}, and $\Delta$ is proportional to the Fermi energy $\epsilon_{\rm F}=\frac{\hbar^2 k_{\rm F}^2}{2m}$,
$\Delta\simeq 0.44 \epsilon_{\rm F}$ \cite{KetterleGap}. This also determines their derivatives with respect to $\rho$.
Similarly, the equation of state measured at $T=0$ is simply $\mu=\xi \epsilon_{\rm F}$, where $\xi\simeq 0.376$ \cite{Zwierlein2012}, and the critical temperature is $T_c\simeq 0.167\epsilon_{\rm F}/k_{\rm B}$ \cite{Zwierlein2012}. For the effective mass of the fermionic excitations and their dispersion relation at non vanishing $k-k_0$, we must rely on results of a dimensional $\epsilon=4-d$ expansion,
$m_*/m\simeq 0.56$ and $\epsilon_k\simeq\Delta+\frac{\hbar^2(k^2-k_0^2)^2}{8m_* k_0^2}$ \cite{Nishida}.  We also trust Anderson's RPA prediction \cite{Anderson,CKS} that the $q=0$ third derivative of the phononic dispersion relation is positive \cite{concavrpa}. The damping rates of phonons with wavenumber $q=mc/2\hbar$ are plotted in Fig.~\ref{fig:gaz}a. The contribution of the three-phonon Landau-Beliaev processes $\phi\leftrightarrow \phi\phi$, here energetically allowed, is dominant; it is computed in the quantum-hydrodynamic approximation where it is independent of the aforementioned third derivative.

The phononic excitation branch becomes concave in the BCS limit $k_{\rm F}a\to 0^-$ \cite{Strinati}. As visible on Fig.~\ref{fig:gaz}b, the phonon-phonon damping (now governed by the Landau-Khalatnikov processes mentioned earlier) is much weaker, and dominates the $\phi-\gamma$ damping only at very low temperatures. At commonly reached temperatures $T> 0.05\epsilon_{\rm F}/k_B$ \cite{Ketterlefroid}, the damping is in fact dominated by {\gilles absorption-emission} $\phi-\gamma$ processes which, unlike in {\yvan liquid} helium, prevail over {\gilles scattering ones} because of {\gilles the smaller value of $\epsilon_{k_*}/\Delta$}. Although the associated quality factors $\omega_\qq/\Gamma_\qq$ may seem impressive, the lifetimes $\Gamma_\qq^{-1}$ of the modes do not exceed one second in a gas of ${}^6$Li with a typical Fermi temperature $T_{\rm F}=1\mu$K, which is shorter than what was observed in a Bose-Einstein condensate \cite{Dalibard}. 
Our predictions, less quantitative than on Fig.~\ref{fig:gaz}a, are based on the BCS approximation for the equation of state and the fermionic excitation dispersion relation $\epsilon_k \simeq \epsilon_k^{\rm BCS}=[(\frac{\hbar^2 k^2}{2m}-\mu)^2+\Delta_{\rm BCS}^2]^{1/2}$ and on the RPA for the $q=0$ third derivative of $\omega_q$ (whose precise value matters here). A cutting remark on Ref.\cite{VincentLiu}: even in the BCS approximation to which it is restricted, we disagree with its expression of $\Gamma_\qq^{\gilles\mbox{\scriptsize a-e}}$.

\noindent{\sl Conclusion --} 
By complementing the {\yvan local density approximation in Ref.\cite{LKhydroquant}} with a systematic low-temperature expansion, we derived the definitive leading order expression of the phonon-roton coupling in liquid helium and we generalized it to the phonon-pair-breaking excitation coupling in Fermi gases. The ever-improving experimental technics in these systems give access to the microscopic parameters determining the coupling and allow for a verification in the near future. Our result also clarifies the regime of temperature and interaction strength in which the purely phononic $\phi\phi\leftrightarrow\phi\phi$ Landau-Khalatnikov sound damping in a superfluid, unobserved to this day, is dominant.

\begin{acknowledgments}
This project received funding from the FWO and the EU H2020 program under the MSC Grant Agreement No. 665501.
\end{acknowledgments}

\appendix
\bigskip
\centerline{\fbox{\bf Addition in the corrected-augmented version}}

\section{Erratum : corrections due to three-body interactions among phonons.}
\label{ann:erratum}

The results on the effective $\phi-\gamma$ scattering amplitude and associated damping rate, based on the roton Hamiltonian (Eqs.~ \eqref{eq:hrot} and \eqref{eq:hrotdevsec}), do not account for the phonon-phonon interaction, which is
described, to lowest order in  $q$, by the Hamiltonian \cite{LKhydroquant}:
\begin{multline}
\hat{H}^{(3)}_{\rm \phi-\phi} =  \sum_{\qq_1,\qq_2,\qq_3} \delta_{\qq_1+\qq_2,\qq_3} \frac{\mathcal{A}^{2\leftrightarrow1}_{\rm \phi-\phi}(\qq_1,\qq_2;\qq_3)}{\mathcal{V}^{1/2}}\\
\times \bb{\hat b_{\qq_1}^\dagger \hat b_{\qq_2}^\dagger \hat b_{\qq_3}+\mbox{h.c.}} 
\end{multline}
with the $\phi+\phi\leftrightarrow \phi$ amplitude \cite{Annalen}:
\begin{multline}
\mathcal{A}^{2\leftrightarrow1}_{\rm \phi-\phi}(\qq_1,\qq_2;\qq_3)=\frac{mc^2}{\rho^{1/2}}\sqrt{\frac{\hbar^3q_1 q_2 q_3}{32 m^3c^3}}\\ 
\times \bb{2\frac{\dd\ln c}{\dd\ln \rho}-1+\frac{\qq_1\cdot\qq_2}{q_1 q_2}+\frac{\qq_1\cdot\qq_3}{q_1 q_3}+\frac{\qq_2\cdot\qq_3}{q_2 q_3}}
\end{multline}
We have used here the Gr\"uneisen parameter $\frac{\dd\ln c}{\dd\ln \rho}=(\rho\mu''/\mu'+1)/2$.
To leading order, there are two diagrams mediated by $\hat{H}^{(3)}_{\rm \phi-\phi}$ missing in Eq.~\eqref{eq:A2eff}:
\begin{widetext}

\hspace{-3mm}\includegraphics[width=\textwidth,clip=]{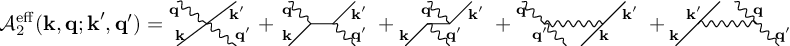}
\begin{multline}
\label{eq:A2efferr}
=\mathcal{A}_2(\kk,\qq;\kk',\qq')
+\frac{\mathcal{A}_1(\kk,\qq;\kk+\qq) \mathcal{A}_1(\kk',\qq';\kk'+\qq')}{\hbar\omega_\qq + \epsilon_\kk-\epsilon_{\kk+\qq}} 
+\frac{\mathcal{A}_1(\kk-\qq',\qq';\kk) \mathcal{A}_1(\kk-\qq',\qq;\kk')}{\epsilon_\kk-\hbar\omega_{\qq'}-\epsilon_{\kk-\qq'}} \\
+\frac{2\mathcal{A}^{2\leftrightarrow1}_{\rm \phi-\phi}(\qq',\qq-\qq';\qq) \mathcal{A}_1(\kk,\qq-\qq';\kk')}{\hbar(\omega_{\qq}-\omega_{\qq-\qq'}-\omega_{\qq'})}+\frac{2\mathcal{A}^{2\leftrightarrow1}_{\rm \phi-\phi}(\qq,\qq'-\qq;\qq') \mathcal{A}_1(\kk',\qq'-\qq;\kk)}{\epsilon_{\kk}-\epsilon_{\kk'}-\hbar\omega_{\qq-\qq'}} 
\end{multline}
The energy denominators of the two new diagrams are equal to
$-\hbar\omega_{\qq-\qq'}$ to leading order (see note [45]). One should thus add $-({\hbar q}/{mc\rho})  ({\rho\Delta'}/2)  (2{\dd\ln c}/{\dd\ln \rho}-1+w) $ to Eq.~\eqref{eq:A2effequiv}:
\begin{multline}
\label{eq:A2effequiv2}
\mathcal{A}_2^{\rm eff}(\kk,\qq;\kk',\qq')  \underset{T\to 0}{\sim} \frac{\hbar q}{mc\rho}
\Bigg\{\frac{1}{2} \rho^2 \Delta'' -\frac{\rho\Delta'}{2} \bb{2\frac{\dd\ln c}{\dd\ln \rho}-1}+ \frac{(\hbar \rho k_0')^2}{2m_*} 
+ \frac{\hbar^2 k_0^2}{2m_*} \\ \times \Bigg\{\!\!\left(\frac{\rho\Delta'}{\hbar c k_0}\right)^2\!\! u u' + \frac{\rho\Delta'}{\hbar c k_0}
\left[(u+u')\left(u u' -\frac{\rho k_0'}{k_0}\right) + \frac{m_* c}{\hbar k_0} w\right]\\
+ \frac{m_* c}{\hbar k_0} (u+u') w + u^2 u'^2 -\frac{\rho k_0'}{k_0} (u^2+u'^2) \Bigg\}\Bigg\}
\end{multline}
\end{widetext}
This also changes the angular integral given in note [47]: $
I/(4\pi)^2=
(\frac{\hbar^2 k_0^2}{2m_*mc^2})^2[\frac{1}{25}-\frac{4\alpha}{15}+\frac{28}{45}\alpha^2+\frac{2\beta^2}{9}
+A(\frac{2}{9}-\frac{4\alpha}{3})+A^2+4\beta B(\frac{1}{15}-\frac{\alpha}{9})
+B^2(\frac{2}{15}-\frac{4\alpha}{9}+\frac{2\alpha^2}{3}+\frac{\beta^2}{3})+\frac{2\beta}{9}B^3+\frac{B^4}{9}]$, 
and the definition of $A=\frac{m_*}{(\hbar k_0)^2}[\rho^2\Delta''+\rho\Delta'(1-2{\dd\ln c}/{\dd\ln \rho})] +\alpha^2$.
Accordingly, the black dashed curves in Figs.~\ref{fig:helium}, \ref{fig:gaz}a and \ref{fig:gaz}b are multiplied respectively by $0.46$, $0.78$ and $0.85$. Note that the Gr\"uneisen parameter also appears in Ref.~\cite{Penco2018}; our coefficients of the $uu'$ and $w$ monomials of $\mathcal{A}_2^{\rm eff}$ are however still in disagreement with that reference.

\section{Microscopic verification of the effective coupling amplitude}
\label{ann:micro}
To check the expression of the effective coupling amplitude
that we obtained through
phonon-roton hydrodynamics (Eq.~\eqref{eq:A2effequiv2}), 
we recompute it within a microscopic approach in the particular case of a
weakly-interacting Bose gas.
In this well-known system, the excitation spectrum takes the Bogoliubov form
\be
\epsilon_k=\sqrt{\frac{\hbar^2 k^2}{2m}\bb{\frac{\hbar^2 k^2}{2m}+2\rho V_k}}
\label{eBogo}
\ee
where $V_k$ is the Fourier transform of the interaction potential. This spectrum describes hydrodynamic
phonons (with $\epsilon_q=\hbar c q +O(q^3)$) provided $V_0>0$ and $\partial_k V_k\vert_{k=0}=0$. 
It has a roton minimum $\Delta=\epsilon_{k_0}$ in $k=k_0$ provided $\partial_k \epsilon_k\vert_{k=k_0}=0$
and $\partial_k^2 \epsilon_k\vert_{k=k_0}=\hbar^2/m_*>0$, which is always possible
with a careful choice of the function $k\mapsto V_k$.

Next, we use the 3- and 4-quasiparticle coupling amplitudes derived from
Bogoliubov theory
(see Eqs.~(E18), (E19) and (E20) of Ref.~\cite{Annalen}), we compute
the effective scattering amplitude in second order perturbation theory as prescribed by Eq.~(105) of \cite{Annalen}
(in which we set $\qq_1=\qq$, $\qq_2=\kk$, $\qq_3=\qq'$ and $\qq_4=\kk'$) and we take the limit of $q$ and $q'$ tending to $0$. 
Since the expected result 
\eqref{eq:A2effequiv2} does not depend on $k$ to leading order, we choose $k=k_0$,
which greatly simplifies the microscopic calculation. As in the hydrodynamic approach,
we then eliminate $\kk'$ using momentum conservation,
and the norm of $\qq'$ using energy conservation (with the difference that one should go 
up to order $q^3$ in the calculation of $q-q'$
because the scattering amplitude of the microscopic model diverges as $1/q$ off-shell).
We get finally:
\begin{widetext}
\begin{multline}
\label{eq:A2effmicro}
\mathcal{A}_{\rm micro}^{\rm eff}(\kk,\qq;\kk',\qq')  \underset{T\to 0}{\sim} \frac{\hbar q}{16\Delta^3 mc\rho}
\Bigg\{ -2E_{k_0}^4+\frac{2\Delta^5}{m_*c^2}uu'+\Delta E_{k_0}^3 \frac{\hbar k_0}{mc}\bbcro{4(u+u')+\frac{\hbar k_0}{m_*c}uu'}\\
-4E_{k_0}^2\Delta^2\bbcro{4(u^2+u'^2)+w-1+\frac{\hbar k_0}{m_* c}uu'(u+u')}+2\Delta^4\bbcro{2w-1+\frac{2\hbar k_0}{m_*c}uu'(u+u')}\\
-2\Delta^3 E_{k_0}\frac{\hbar k_0}{mc}\bbcro{2(u+u')+\frac{\hbar k_0}{m_*c}uu'}
+8\Delta^3 mc^2\frac{\hbar k_0}{mc}\bbcro{w(u+u')+\frac{\hbar k_0}{m_*c}u^2u'^2} \Bigg\}
\end{multline}
\end{widetext}
where $E_{k_0}=\hbar^2 k_0^2/2m$. This result coincides with the hydrodynamic expression \eqref{eq:A2effequiv2}
specialized using the Bogoliubov dispersion relation \eqref{eBogo} to express $k_0'$, $\Delta'$ and $\Delta''$ in terms of $k_0$, $\Delta$ and $m_*$:
\be
k_0'\!=\!\frac{m_* k_0 E_{k_0}}{m\rho\Delta}\,,\ \Delta'\!=\!\frac{\Delta^2-E_{k_0}^2}{2\rho\Delta}\,,\  \Delta''\!=\!-\frac{\hbar^2 k_0 k_0'E_{k_0}}{m\rho\Delta}-\frac{\Delta'^2}{\Delta}
\ee
and the Bogoliubov equation of state $\mu=\rho V_0$
to obtain $\mu'=V_0$ and $\mu''=0$.

\centerline{\fbox{\bf End of addition}}


\begin{thebibliography}{0}
\expandafter\ifx\csname natexlab\endcsname\relax\def\natexlab#1{#1}\fi
\expandafter\ifx\csname bibnamefont\endcsname\relax
  \def\bibnamefont#1{#1}\fi
\expandafter\ifx\csname bibfnamefont\endcsname\relax
  \def\bibfnamefont#1{#1}\fi
\expandafter\ifx\csname citenamefont\endcsname\relax
  \def\citenamefont#1{#1}\fi
\expandafter\ifx\csname url\endcsname\relax
  \def\url#1{\texttt{#1}}\fi
\expandafter\ifx\csname urlprefix\endcsname\relax\def\urlprefix{URL }\fi
\providecommand{\bibinfo}[2]{#2}
\providecommand{\eprint}[2][]{\url{#2}}

\end{thebibliography}


\begin{thebibliography}{0}%
\makeatletter
\providecommand \@ifxundefined [1]{%
 \@ifx{#1\undefined}
}%
\providecommand \@ifnum [1]{%
 \ifnum #1\expandafter \@firstoftwo
 \else \expandafter \@secondoftwo
 \fi
}%
\providecommand \@ifx [1]{%
 \ifx #1\expandafter \@firstoftwo
 \else \expandafter \@secondoftwo
 \fi
}%
\providecommand \natexlab [1]{#1}%
\providecommand \enquote  [1]{``#1''}%
\providecommand \bibnamefont  [1]{#1}%
\providecommand \bibfnamefont [1]{#1}%
\providecommand \citenamefont [1]{#1}%
\providecommand \href@noop [0]{\@secondoftwo}%
\providecommand \href [0]{\begingroup \@sanitize@url \@href}%
\providecommand \@href[1]{\@@startlink{#1}\@@href}%
\providecommand \@@href[1]{\endgroup#1\@@endlink}%
\providecommand \@sanitize@url [0]{\catcode `\\12\catcode `\$12\catcode
  `\&12\catcode `\#12\catcode `\^12\catcode `\_12\catcode `\%12\relax}%
\providecommand \@@startlink[1]{}%
\providecommand \@@endlink[0]{}%
\providecommand \url  [0]{\begingroup\@sanitize@url \@url }%
\providecommand \@url [1]{\endgroup\@href {#1}{\urlprefix }}%
\providecommand \urlprefix  [0]{URL }%
\providecommand \Eprint [0]{\href }%
\providecommand \doibase [0]{http://dx.doi.org/}%
\providecommand \selectlanguage [0]{\@gobble}%
\providecommand \bibinfo  [0]{\@secondoftwo}%
\providecommand \bibfield  [0]{\@secondoftwo}%
\providecommand \translation [1]{[#1]}%
\providecommand \BibitemOpen [0]{}%
\providecommand \bibitemStop [0]{}%
\providecommand \bibitemNoStop [0]{.\EOS\space}%
\providecommand \EOS [0]{\spacefactor3000\relax}%
\providecommand \BibitemShut  [1]{\csname bibitem#1\endcsname}%
\let\auto@bib@innerbib\@empty
\end{thebibliography}%


\begin{thebibliography}{99}

\bibitem{LKhydroquant}
L. Landau, I. Khalatnikov, 
\g{Teoriya vyazkosti Geliya-II},
Zh. Eksp. Teor. Fiz. {\bf 19}, 637 (1949) {\yvan [English translation in {\sl Collected papers of L.D. Landau},  chapter 69, pp.494-510, edited by D. ter Haar
(Pergamon, New York, 1965)].}

\bibitem{Chernikova}
I. M. Khalatnikov, D.M. Chernikova, \g{Relaxation phenomena in superfluid Helium},
Zh. Eksp. Teor. Fiz. {\bf 49}, 1957 (1965) [JETP {\bf 22}, 1336 (1966)].

\bibitem{Fok}
B. F\aa k, T. Keller, M. E. Zhitomirsky, A. L. Chernyshev,
\g{Roton-phonon interaction in superfluid ${}^4$He},
Phys. Rev. Lett. {\bf 109}, 155305 (2012).

\bibitem{boite}
A. L. Gaunt, T. F. Schmidutz, I. Gotlibovych, R. P. Smith, Z. Hadzibabic,
\g{Bose-Einstein condensation of atoms in a uniform potential},
Phys. Rev. Lett. {\bf 110}, 200406 (2013).

\bibitem{boite_fermions}
B. Mukherjee, Zhenjie Yan, P. B. Patel, Z. Hadzibabic, T. Yefsah, J. Struck, M. W. Zwierlein,
\g{Homogeneous Atomic Fermi Gases},
Phys. Rev. Lett. {\bf 118}, 123401 (2017).

\bibitem{EPL}
H. Kurkjian, Y. Castin, A. Sinatra,
\g{Landau-Khalatnikov phonon damping in strongly interacting Fermi gases},
EPL {\bf 116}, 40002 (2016).

\bibitem{MZP}
{\gilles Martin Zwierlein, private communication (September 2017).}

\bibitem{CCT}
{\yvan C. Cohen-Tannoudji, \g{Atomic motion in laser light}, \S 2.3, in {\sl Proceedings of the Les Houches Summer School}, session LIII, 
edited by J. Dalibard, J.-M. Raimond, J.  Zinn-Justin (North-Holland, Amsterdam, 1992).}

\bibitem{Thomas}
{\yvan L. H. Thomas, \g{The calculation of atomic fields},  Proc. Cambridge Phil. Soc. {\bf 23}, 542 (1927).}

\bibitem{Fermi}
{\yvan E. Fermi, \g{Un metodo statistico per la determinazione di alcune priopriet\`a  dell'atomo}, Rend. Accad. Naz. Lincei {\bf 6}, 602 (1927)
[\g{A statistical method to evaluate some properties of the atom}] and in {\sl Collected papers, Note e memorie of Enrico Fermi, volume I},
edited by E. Amaldi {\sl et al.} (The University of Chicago Press, Chicago, 1962).}

\bibitem{Beliaev}
S. T. Beliaev, \g{Energy-Spectrum of a Non-ideal Bose Gas}, Zh. Eksp. Teor. Fiz. {\bf 34}, 433 (1958) [JETP {\bf 7}, 299 (1958)].

\bibitem{PitaevskiiStringari}
L. P. Pitaevskii, S. Stringari, \g{Landau damping in dilute Bose gases}, Phys. Lett. A {\bf 235}, 398 (1997).

\bibitem{Giorgini}
S. Giorgini, \g{Damping in dilute Bose gases: A mean-field approach}, Phys. Rev. A {\bf 57}, 2949 (1998).

\bibitem{mesuresHe}
B. M. Abraham, Y. Eckstein, J. B. Ketterson, M. Kuchnir, J. Vignos, \g{Sound Propagation in Liquid ${}^{4}$He}, Phys. Rev. {\bf 181}, 347 (1969).

\bibitem{mesuresBose} N. Katz, J. Steinhauer, R. Ozeri, N. Davidson,
\g{Beliaev Damping of Quasiparticles in a Bose-Einstein Condensate}, Phys. Rev. Lett. {\bf 89}, 220401 (2002).

\bibitem{Annalen}
H. Kurkjian, Y. Castin, A. Sinatra,
\g{Three-phonon and four phonon interaction processes in a pair-condensed Fermi gas},
Ann. Phys. (Berlin) {\yvan {\bf 529}}, 1600352 (2017).

\bibitem{Lockerbie1974}
N. A. Lockerbie, A. F. G. Wyatt, R. A. Sherlock,
\g{Measurement of the group velocity of 93 GHz phonons in liquid ${}^4$He},
Solid State Communications {\bf 15}, 567 (1974).

\bibitem{Dietsche1978}
W. Dietsche,
\g{Superconducting Al-PbBi tunnel junction as a phonon spectrometer},
Phys. Rev. Lett. {\bf 40}, 786 (1978).

\bibitem{Berberich1976}
P. Berberich, P. Leiderer, S. Hunklinger,
\g{Investigation of the lifetime of longitudinal phonons at GHz frequencies in liquid and solid  ${}^4$He},
Journal of Low Temperature Physics {\bf 22}, 61 (1976).

\bibitem{FosterPRB30_2595}
D. Rugar, J. S. Foster,
\g{Accurate measurement of low-energy phonon dispersion in liquid ${}^4$He},
Phys. Rev. B {\bf 30}, 2595 (1984).

\bibitem{SvensonPRL29_1148}
E. C. Swenson, A. D. B. Woods, P. Martel,
\g{Phonon dispersion in liquid Helium under pressure},
Phys. Rev. Lett. {\bf 29}, 1148 (1972).

\bibitem{Gibbs1999} 
M. R. Gibbs, K. H. Andersen, W. G. Stirling, H. Schober,
\g{The collective excitations of normal and superfluid ${}^4$He: the dependence on pressure and temperature},
J. Phys. Condens. Matter {\bf 11}, 603 (1999).

\bibitem{Marris2002}
H. J. Maris, D. O. Edwards,
\g{Thermodynamic properties of superfluid ${}^4$He at negative pressure},
Journal of Low Temperature Physics {\bf 129}, 1 (2002).

\bibitem{Thomas2002}
K. M. O'Hara, S. L. Hemmer, M. E. Gehm, S. R. Granade, J. E. Thomas, 
\g{Observation of a strongly interacting degenerate Fermi gas of atoms},
Science {\bf 298}, 2179 (2002).

\bibitem{Salomon2003}
T. Bourdel, J. Cubizolles, L. Khaykovich, K. M. Magalh\~aes, S. J. J. M. F. Kokkelmans, G. V. Shlyapnikov, C. Salomon,
\g{Measurement of the interaction energy near a Feshbach resonance in a $^{6}\mathrm{L}\mathrm{i}$ Fermi gas},
Phys. Rev. Lett. {\bf 91}, 020402 (2003).

\bibitem{Grimm2004b}
M. Bartenstein, A. Altmeyer, S. Riedl, S. Jochim, C. Chin, J. H. Denschlag, R. Grimm,
\g{Collective excitations of a degenerate gas at the BEC-BCS crossover},
Phys. Rev. Lett. {\bf 92}, 203201 (2004).

\bibitem{Ketterle2004}
M. W. Zwierlein, C. A. Stan, C. H. Schunck, S. M. F. Raupach, A. J. Kerman, W. Ketterle,
\g{Condensation of pairs of fermionic atoms near a Feshbach resonance},
Phys. Rev. Lett. {\bf 92}, 120403 (2004).

\bibitem{Salomon2010}
S. Nascimb{\`e}ne, N. Navon, K. J. Jiang, F. Chevy, C. Salomon, 
\g{Exploring the thermodynamics of a universal Fermi gas},
Nature {\bf 463}, 1057 (2010).

\bibitem{Zwierlein2012}
M. J. H. Ku, A. T. Sommer, L. W. Cheuk, M. W. Zwierlein,
\g{Revealing the superfluid lambda transition in the universal thermodynamics of a unitary Fermi gas},
Science {\bf 335}, 563 (2012).

\bibitem{KetterleGap}
A. Schirotzek, Y. Shin, C. H. Schunck, W. Ketterle, 
\g{Determination of the superfluid gap in atomic Fermi gases by quasiparticle spectroscopy},
Phys. Rev. Lett.  {\bf 101}, 140403 (2008).

\bibitem{JasonHo}
Tin-Lun Ho, \g{Universal thermodynamics of degenerate quantum gases in the unitarity limit},
Phys. Rev. Lett. {\bf 92}, 090402 (2004).

\bibitem{Zwerger}
T. Enss, R. Haussmann, W. Zwerger,
\g{Viscosity and scale invariance in the unitary Fermi gas},
Annals of Physics  {\bf 326}, 770 (2011).

\bibitem{Zwergerlivre}
Y. Castin, F. Werner, \g{The Unitary Gas and its Symmetry Properties},
in {\sl BCS-BEC Crossover and the Unitary Fermi gas},
Springer Lecture Notes in Physics, W. Zwerger ed. (Springer, Berlin, 2011).

\bibitem{Nishida}
Y. Nishida, D. T. Son,
\g{$\epsilon$ Expansion for a Fermi gas at infinite scattering length},
Phys. Rev. Lett. {\bf 97}, 050403 (2006).

\bibitem{Anderson}
P.W. Anderson,
\g{Random-phase approximation in the theory of superconductivity},
Phys. Rev. {\bf 112}, 1900 (1958).

\bibitem{CKS}
R. Combescot, M. Y. Kagan, S. Stringari,
\g{Collective mode of homogeneous superfluid Fermi gases in the BEC-BCS crossover},
Phys. Rev. A  {\bf 74}, 042717 (2006).

\bibitem{concavrpa}
H. Kurkjian, Y. Castin, A. Sinatra,
\g{Concavity of the collective excitation branch of a Fermi gas in the BEC-BCS crossover},
Phys. Rev. A {\bf 93}, 013623 (2016).

\bibitem{Strinati}
M. Marini, F. Pistolesi, G. C. Strinati,
\g{Evolution from BCS superconductivity to Bose condensation: analytic results for the crossover in three dimensions},
European Physical Journal B {\bf 1}, 151 (1998).

\bibitem{Ketterlefroid}
Z. Hadzibabic, S. Gupta, C. A. Stan, C. H. Schunck, M. W. Zwierlein, K. Dieckmann, W. Ketterle,
\g{Fiftyfold Improvement in the Number of Quantum Degenerate Fermionic Atoms},
Phys. Rev. Lett. {\bf 91}, 160401 (2003).

\bibitem{Dalibard}
F. Chevy, V. Bretin, P. Rosenbusch, K. W. Madison, J. Dalibard,
\g{Transverse Breathing Mode of an Elongated Bose-Einstein Condensate},
Phys. Rev. Lett. {\bf 88}, 250402 (2002).

\bibitem{VincentLiu}
Z. Zhang, W. Vincent Liu,
\g{Finite-temperature damping of collective modes of a BCS-BEC crossover superfluid},
Phys. Rev. A {\bf 83}, 023617 (2011).

\bibitem{erratum}
Y. Castin, A. Sinatra, H. Kurkjian, \g{Erratum: Landau Phonon-Roton Theory Revisited for Superfluid ${}^4$He and Fermi Gases [PRL {\bf 119},260402]}, to appear in Phys. Rev. Lett. (2019).

\bibitem{Penco2018}
A. Nicolis, R. Penco,
\g{Mutual interactions of phonons, rotons, and gravity},
Phys. Rev. B, {\bf 97}, 134516 (2018).

\end{thebibliography}
\end{document}